\documentclass[11pt,english]{oxarticle}
\pdfoutput=1

\usepackage{graphicx, float, array, xspace, amscd, amsmath, amsthm, amssymb, latexsym, bbm, longtable, tikz-cd, mathtools, caption, setspace, multirow,colortbl}
\usepackage[hyperfootnotes=false, linktocpage=true, colorlinks, citecolor=blue, linkcolor=blue, urlcolor=brown]{hyperref}
\usepackage[all,pdf]{xy}\SelectTips {cm}{}
\usepackage[bottom]{footmisc}
\usepackage[sort&compress, comma, square, numbers]{natbib}
\usepackage[normalem]{ulem}
\usepackage{html}
\usepackage{navigator}

\DeclareFontFamily{OT1}{pzc}{}
\DeclareFontShape{OT1}{pzc}{m}{it}{<-> s * [1.200] pzcmi7t}{}
\DeclareMathAlphabet{\mathpzc}{OT1}{pzc}{m}{it}

\newcommand{\IA}{\mathbb{A}}

\newcommand{\ID}{\mathbb{D}}

\newcommand{\IP}{\mathbb{P}}
\newcommand{\IQ}{\mathbb{Q}}

\newcommand{\IS}{\mathbb{S}}

\newcommand{\IZ}{\mathbb{Z}}

\font\eightrm=cmr8 at 8pt

\DeclareFontFamily{U}{wncy}{}
\DeclareFontShape{U}{wncy}{m}{n}{<->wncyr10}{}
\DeclareSymbolFont{mcy}{U}{wncy}{m}{n}
\DeclareMathSymbol{\sha}{\mathord}{mcy}{"58}

\newcommand{\varstr}[2]{\vrule height #1 depth #2 width0pt}

\newcommand{\place}[3]{\vbox to0pt{\kern-\parskip\kern-7pt
                             \kern-#2truein\hbox{\kern#1truein #3}
                             \vss}\nointerlineskip}

\newcommand{\beq}{\begin{equation}}
\newcommand{\eeq}{\end{equation}}
\newcommand{\beqnn}{\begin{equation*}}
\newcommand{\eeqnn}{\end{equation*}}

\newcommand{\+}{\hphantom{-}}

\def\fnote#1#2{\begingroup\def\thefootnote{#1}\footnote{#2}
     \addtocounter{footnote}{-1}\endgroup}

\newcommand{\cicy}[2]{\begin{matrix} #1\end{matrix}\!\left[\begin{matrix}#2 \end{matrix}\right]}

\newcommand{\quotient}[1]{_{\hskip-2pt\lower1pt\hbox{$/$}\lower2pt\hbox{\hskip-1pt$#1$}}}

\numberwithin{equation}{section}

\newlength{\xtrawidth}
\newlength{\xtraheight}
\proofmodefalse
\newlength{\myht}
\newlength{\mydp}
\newlength{\mywd}
\newsavebox{\mybox}
\def\str{\vrule height14pt depth8pt width0pt}
\begin{document}
\pagestyle{empty}
\begin{center}
\null\vskip0.3in
{\LARGE\bf Discrete Symmetries of Complete Intersection Calabi-Yau Manifolds\\[1cm]}
{Andre Lukas and Challenger Mishra\\[6mm]}
{\it Rudolf Peierls Centre for Theoretical Physics\hphantom{$^3$}\\
University of Oxford\\
1 Keble Road, 
Oxford OX1 3NP, UK\\[2cm]}
\end{center}

\begin{abstract}\noindent
In this paper, we classify non-freely acting discrete symmetries of complete intersection Calabi-Yau manifolds and their quotients by freely-acting symmetries. These non-freely acting symmetries can appear as symmetries of low-energy theories resulting from string compactifications on these Calabi-Yau manifolds, particularly in the context of the heterotic string. Hence, our results are relevant for four-dimensional model building with discrete symmetries and they give an indication which symmetries of this kind can be expected from string theory. For the 1695 known quotients of complete intersection manifolds by freely-acting discrete symmetries,  non-freely-acting, generic symmetries arise in 381 cases and are, therefore, a relatively common feature of these manifolds. We find that 9 different discrete groups appear, ranging in group order from 2 to 18, and that both regular symmetries and R-symmetries are possible. 
\end{abstract}

\fnote{}{lukas@physics.ox.ac.uk} 
\fnote{}{challenger.mishra@physics.ox.ac.uk}
\newpage
\begingroup
\baselineskip=14pt
\tableofcontents
\endgroup
\newpage
\setcounter{page}{1}
\pagestyle{plain}
\section{Introduction}
\vskip-6pt
Finite symmetries are a widely-used tool in particle physics model building, particularly in the context of models for fermion masses and as a way to forbid unwanted operators such as those inducing fast proton decay. Usually, such symmetries are required to be discrete gauge symmetries~\cite{Ibanez:1991pr} but no further theoretical constraints are placed on them at this level. Within the framework of supersymmetric model building both regular symmetries and R-symmetries are considered. 

It is natural to ask how such discrete symmetries can arise in string theory and which specific groups can be obtained from string compactifications. Discrete symmetries in four-dimensional string models arise from remnants of local string symmetries which remain unbroken by the compactification. In particular, this means that discrete symmetries from string theory are always discrete gauge symmetries. Specifically, such symmetries can arise as symmetries of  the compactification manifolds and studying these manifolds and their properties is a good way to identify discrete symmetries from string theory. The main purpose of the present paper is to study a certain class of Calabi-Yau (CY) three-folds from this point of view. Genuine Calabi-Yau manifolds (that is manifolds with holonomy group $SU(3)$) do not have continuous symmetries but it is well-known that discrete symmetries can arise.

Most of the literature on CY symmetries to date has been concerned with freely-acting symmetries. For complete intersection CY manifolds (CICY manifolds)~\cite{Candelas:1987kf}, considerable work has been carried out to find freely-acting symmetries~\cite{Candelas:1987du,Candelas:2008wb,Candelas:2010ve}, culminating in the classification of Ref.~\cite{Braun:2010vc} which provides all freely-acting symmetries of CICY manifolds which descend from linear actions on the projective ambient spaces. There has been less work on freely-acting discrete symmetries of CY manifolds defined as hyper-surfaces in toric four-folds. In Ref.~\cite{Batyrev:2005jc}, all toric freely-acting symmetries have been classified for those manifolds and have been found to exist for only 16 of the about half a billion reflexive polytopes. A first step towards a systematic classification has recently been made in Ref.~\cite{Braun:2017juz}. 
Over the past few years there has also been considerable work on discrete symmetries in the context of type II string compactifications~\cite{BerasaluceGonzalez:2011wy}--\cite{Berasaluce-Gonzalez:2013bba}.

Freely-acting symmetries of CY manifolds are useful to construct new CY manifolds with a non-trivial first fundamental group by forming quotients and for CICY manifolds much work in this direction has been carried out~\cite{Candelas:2008wb,Candelas:2010ve,Candelas:2015amz,Candelas:2016fdy,Constantin:2016xlj}. Such quotient CY manifolds are a vital ingredient for compactifications of the heterotic string, where the standard model building paradigm demands the presence of a Wilson line and, hence, a non-trivial first fundamental group of the CY manifold. Of course, freely-acting CY symmetries which have been divided out no longer give rise to discrete low-energy symmetries. Those arise from CY symmetries, both freely and non-freely acting, which have not been divided out. 

At first sight, the experience with freely-acting CY symmetries is discouraging in view of generating low-energy discrete symmetries from string theory. CY manifolds come with moduli spaces and in all known examples freely-acting symmetries only appear at special lower-dimensional sub-loci in moduli space. For the corresponding low-energy theory, this means that the discrete symmetry is generically not visible and will only appear if the moduli fields are tuned to the relevant sub-locus. For example, the quintic in $\mathbb{P}^4$ has a complex structure moduli space of dimension $101$ and a freely-acting $\mathbb{Z}_5\times\mathbb{Z}_5$ symmetry which appears only at a $5$-dimensional sub-locus on this moduli space. This seems to suggest that discrete symmetries in four-dimensional string models, at least insofar as they originate from symmetries of the compactification manifold, are typically quite non-generic and are unlikely to play a major role in phenomenology. One of the main points of the present paper is that this statement does not apply to certain classes of CY manifolds. This means that, contrary to expectation, low-energy symmetries which descend from these manifolds can be generic and phenomenologically relevant. 

We will be working in the context of CICY manifolds, $X$, taken from the original list in Ref.~\cite{Candelas:1987kf}, and their freely-acting symmetries, $G_\text{f}$, as classified in Ref.~\cite{Braun:2010vc}. More precisely, our starting point will be all pairs $(X,G_\text{f})$ of these objects  and the associated quotient CY manifolds $Y=X/G_\text{f}$, a dataset of $1695$ manifolds. We recall that such quotient CY manifolds are the preferred starting point for heterotic string compactifications and CICY quotients have indeed been used extensively for heterotic model building~\cite{Greene:1986bm,Greene:1986jb,Candelas:1987du,Anderson:2009mh,Anderson:2011ns,Anderson:2012yf,Braun:2011ni}. In this context, we are asking and answering the following questions. Which of the CICY quotients $Y$ have a symmetry, freely or non-freely acting, at every point in their complex structure moduli space, which finite groups can arise in this way and how do these groups act on the manifold? As in the classification of freely-acting groups in Ref.~\cite{Braun:2010vc}, we will restrict ourselves to symmetries which act linearly on the projective ambient space coordinates. It is already known, from the work in Ref.~\cite{Witten:1985xc}, that such generic symmetries for CICY quotients do exist. In this paper, it was shown that the quintic quotient by the freely-acting $\mathbb{Z}_5\times\mathbb{Z}_5$ symmetry has a generic, non-freely acting $\mathbb{Z}_2$ symmetry. However, it is not clear how common the appearance of such generic symmetries is among CICY quotients (or, indeed, more generally, among CY quotients).

We will search for these symmetries by analysing the symmetries of the up-stairs CICY $X$, generalising a method proposed in Ref.~\cite{Witten:1985xc}. CICY manifolds are defined as the common zero locus of homogeneous polynomials $p_a$, where $a=1,\ldots ,K$, in an ambient space of the form ${\cal A}=\mathbb{P}^{n_1}\times\cdots\times\mathbb{P}^{n_m}$. A necessary condition for a symmetry transformation of the ambient space symmetry group $G$ to descend to the CICY quotient $Y=X/G_\text{f}$ is that it normalises the freely-acting group $G_\text{f}$. The first step is, therefore, to impose this condition and find the normaliser group $N_G(G_\text{f})$ of $G_\text{f}$. Next, we will find the sub-group $N^\star_G(G_\text{f})\subset N_G(G_\text{f})$ which leaves all $G_\text{f}$ invariant CICY manifolds $X$ invariant. The desired symmetry group, $G_\text{Y}$, of the CICY quotient $Y$ is then given by $G_\text{Y}=N^\star_G(G_\text{f})/G_\text{f}$. In this way, we determine all symmetry groups of the CICY quotients. Using a standard method, for example explained in Ref.~\cite{gsw2}, we can check the transformation of the holomorphic $(3,0)$ form $\Omega$  to determine whether these are regular symmetries or R-symmetries. 

Our results can be summarised as follows. From the $1695$ CICY quotients which can be constructed using the original CICY list~\cite{Candelas:1987kf} and the classification of freely-acting symmetries in Ref.~\cite{Braun:2010vc}, we find $381$ have a generic symmetry of the kind described above. Of these, $113$ quotients are found to have an R-symmetry while the others have a regular symmetry only.  The group $\IZ_2$ is the most common one and it turns out that $8$ other groups can appear, namely
{$\IZ_3$}, {$\IZ_4$}, {$\IZ_2^2$}, {$\IZ_2^3$}, {$\ID_8$}, {$\IZ_2^4$}, {$\IZ_2{\times}\ID_8$} and {$(\IZ_3{\times}\IZ_3){\rtimes}\IZ_2$}. From these all but {$\IZ_2^4$} can appear as a regular symmetries and only $\IZ_2^n$, where $n=1,2,3,4$, allows for an R-symmetry. A detailed account of the results can be found in Section~\ref{sec:results} and the frequency with which the above symmetry groups appear is provided in Table~\ref{tab:NewGroupLis}. In summary, this result means that the appearance of symmetries which are present everywhere in moduli space is, perhaps surprisingly, common among CICY quotients. 

The outline of the paper is as follows. In the following section, we set up the notation and describe the general method we will be using for the classification in detail. In Section~\ref{sec:ex}, this method will be illustrated by a number of simple examples. Our results are presented in Section~\ref{sec:results} and we conclude in Section~\ref{sec:conclusion}. Appendix~\ref{Appendix:AutGrs} provides a table of automorphism groups which enter the classification algorithm and Appendix~\ref{app:results} contains a detailed account of our main result, a table which lists the symmetry groups for all $1695$ CICY quotients.

\section{The method}\label{sec:method}
\vskip-6pt
We begin by setting up the basic framework and our notation. The basic arena are ambient spaces of the form
\begin{equation}
 {\cal A}=\mathbb{P}^{n_1}\times\cdots\times\mathbb{P}^{n_m}\; ,
\end{equation}
which consist of a product of $m$ projective factors, each with dimension $n_i$, and with a total dimension $d=\sum_{i=1}^mn_i$. The homogeneous coordinates for the $t^{\rm th}$ projective factor are denoted by $x_{i}^{\alpha}$, where $\alpha=0,1,\ldots ,n_i$ or, alternatively, by ${\bf x}_i=(x_{i}^0,\ldots ,x_{i}^{n_i})^T$ and ${\bf x}=({\bf x}_1,\ldots ,{\bf x}_m)^T$ refers to all homogeneous coordinates of ${\cal A}$.

The linear automorphism group $G$ of ${\cal A}$ is given by
\begin{equation}
 G=S\ltimes\left({\rm PGL}(n_1+1,\mathbb{C})\times\cdots\times{\rm PGL}(n_m+1,\mathbb{C})\right)\; , \label{Gdef}
\end{equation} 
 where $S$ is the sub-group of the symmetric group $S_m$ which permutes projective factors in ${\cal A}$ with the same dimension.
 
The CICY manifolds $X\subset{\cal A}$ is defined as the common zero locus of polynomials $p_a$, where $a=1,\ldots ,K$, each with multi-degree ${\bf q}_a=(q^1_a,\ldots ,q^m_a)^T$. Since we are interested in three-folds, $K=d-3$ such polynomials are required and in order to obtain CY manifolds the conditions $\sum_{a=1}^Kq_a^i=n_i+1$ have to be imposed for all $i=1,\ldots ,m$. The information on dimensions and degrees is usually summarised by a configuration matrix
\begin{equation}
 X\;=\;\cicy{\IP^{n_1}\\\vdots\\ \IP^{n_m}}{q_1^1&\cdots&q_K^1\\ \vdots&\ddots&\vdots\\ q_1^m&\cdots&q_K^m}^{h^{1,1}(X),h^{2,1}(X)}_{\chi(X)} \; ,
\end{equation}
where the Hodge numbers $h^{1,1}(X)$ and $h^{2,1}(X)$ have been attached as a superscript and the Euler number $\chi(X)$ as a subscript. 
A configuration matrix represents an entire family of CICY manifolds, parametrised by the complex structure, which is encoded in the arbitrary coefficients which enter the defining polynomials $p_a$.

Next, we consider a freely-acting symmetry $G_\text{f}\subset G$, taken from the classification of Ref.~\cite{Braun:2010vc}, and assume the polynomials $p_a$ have been specialised such that $G_\text{f}$ is indeed a symmetry of $X$. We can then form the CICY quotient $Y=X/G_\text{f}$ and it is the symmetry of this quotient we are primarily interested in. More precisely, we would like to determine the symmetry of $Y$ which is present everywhere in its complex structure moduli space. 

This will be done by working in the ``upstairs" picture, that is, by studying the symmetry of $G_\text{f}$ invariant CICY manifolds $X$, generalising a method proposed in Ref.~\cite{Witten:1985xc}. As mentioned earlier, we will focus on symmetries which are linearly realised on the ambient space ${\cal A}$ and, hence, we will be interested in symmetries contained in the ambient space symmetry group $G$ in Eq.~\eqref{Gdef}. For a symmetry $g\in G$ of a $G_\text{f}$ invariant CICY $X$ to descend to the quotient $Y$ it needs to normalise $G_\text{f}$, that is, it needs to satisfy $g\,G_\text{f}=G_\text{f}\,g$. Hence, our first step will be to determine the normaliser group
\begin{equation}
 N_G(G_\text{f}):=\{g\in G\,|\, g\,G_\text{f}=G_\text{f}\, g\}\; . \label{normdef}
\end{equation} 
Computing the normaliser directly, by solving the defining relation
\begin{equation}
 g\,G_\text{f}=G_\text{f}\, g\; , \label{normcond}
\end{equation} 
can be computationally intense since it involves the entire group $G_\text{f}$. In order to circumvent this problem, it is useful to note that every $g\in N_G(G_\text{f})$ defines an automorphism $\psi_g:G_\text{f}\rightarrow G_\text{f}$ via
\begin{equation}
 \psi_g(g_\text{f})=g\, g_\text{f}\, g^{-1}\; . \label{psidef}
\end{equation} 
The centraliser of $G_\text{f}$ in $G$ can be expressed as $C_G(G_\text{f})={\rm Ker}(g\rightarrow \psi_g)$. The normaliser can now be expressed in terms of the automorphism group ${\rm Aut}(G_\text{f})$ as
\begin{equation}
 N_G(G_\text{f})=\{g\in G\,|\, \exists\,\psi\in {\rm Aut}(G_\text{f})\,:\, g\,g_\text{f} g^{-1} =\psi(g_\text{f})\quad \forall g_\text{f}\in G_\text{f}\}\; .
\end{equation} 
Since the automorphism group ${\rm Aut}(G_\text{f})$ can be computed by purely group-theoretical methods this provides a more practical way of computing the normaliser. For each given automorphism $\psi\in {\rm Aut}(G_\text{f})$, we find all symmetries $g\in G$ which satisfy
\begin{equation}
 g\,g_\text{f} g^{-1} =\psi(g_\text{f})
\label{auteqn}\end{equation}
for all $g_\text{f}\in G_\text{f}$ and the normaliser consists of all $g$ found in this way for all automorphisms $\psi$. The centraliser $C_G(G_\text{f})$ can be obtained by solving Eq.~\eqref{auteqn} for $\psi={\rm id}$.

In practice, we compile a list of generators of all sub-groups of ${\rm Aut}(G_\text{f})$ (and also add $\psi={\rm id}$ to obtained the centraliser) and then solve Eq.~\eqref{psidef} for this list of automorphisms $\psi$. The solutions obtained in this way then provide a set of generators for the normaliser $N_G(G_\text{f})$.

This normaliser group is frequently an infinite group and can have continuous parts. However, we still have to impose invariance of all $G_\text{f}$ invariant manifolds $X$ and, as CY manifolds do not have continuous symmetries, this will select a discrete sub-group $N^\star_G(G_\text{f})\subset N_G(G_\text{f})$ of transformations which leave all such $X$ invariant and normalise the freely-acting group $G_\text{f}$. Our second step, in order to determine $N^\star_G(G_\text{f})$, is, therefore, to find all $g\in N_G(G_\text{f})$ for which we can find a permutation $\rho(g)\in S_K$ such that
\begin{equation}
 {\bf p}({\bf x})=\rho(g)\,{\bf p}(g^{-1}{\bf x})\; , \label{pinv}
\end{equation}
where ${\bf p}=(p_1,\ldots ,p_K)^T$ is any vector of $G_\text{f}$ invariant defining polynomials. We also introduce the analogous sub-group $C^\star_G(G_\text{f})\subset C_G(G_\text{f})$ of the centraliser. 

Finally, the symmetry group $G_\text{Y}$ of the CICY quotient $Y$ is then found by dividing out $G_\text{f}$, that is
\begin{equation}
 G_\text{Y}=N^\star_G(G_\text{f})/G_\text{f}\; .
\end{equation} 

We would also like to decide which $g\in N^\star_G(G_\text{f})$ correspond to regular symmetries and which correspond to R-symmetries and for this purpose we should introduce the holomorphic $(3,0)$ form $\Omega$ on $X$ and its counterpart $\widehat{\Omega}$ on the ambient space ${\cal A}$. The latter can be defined implicitly by the relations
\beq
\begin{aligned}\label{RSymmForm}
\widehat{\Omega}~\wedge dp^1~\wedge\dots&\wedge dp^K = \mu\; ,\\
\hskip-10pt\text{where~~}\mu=\mu_1 \wedge \dots \wedge \mu_m \text{~~and~~}
\mu_j&=\frac{1}{n_j!}\epsilon_{\beta_0 \beta_1 \dots \beta_{n_j}} x_j ^{\beta_0}  dx_j^{\beta_1} \wedge \dots \wedge dx_j^{\beta_{n_j}}\; .
\end{aligned}
\eeq
The $(3,0)$ form on $X$ is then given by the restriction $\Omega=\widehat{\Omega}|_X$. From a standard argument, see, for example Ref.~\cite{gsw2}, symmetries which leave $\Omega$ invariant are regular symmetries and those which transform $\Omega$ into a non-trivial multiple of itself are R-symmetries. Since all our symmetries descend from the ambient space it is, in fact, sufficient to check this transformation property for $\widehat{\Omega}$. In other words, we would like to compute the numbers $F(g)$ in
\begin{equation}
 g^\star\,\widehat{\Omega}=F(g)\widehat{\Omega}\; .
\end{equation} 
A quick inspection of Eqs.~\eqref{RSymmForm} shows that they are given by the simple formula
\begin{equation}
 F(g)=\frac{{\rm det}(g)\,{\rm det}(\rho(g))}{{\rm det}(\pi(g))}\; , \label{Fdef}
\end{equation} 
where $\rho(g)$ is the permutation of polynomials which appears in Eq.~\eqref{pinv} and $\pi(g)\in S$ is the permutation part of $g$ which corresponds to the first factor in the definition~\eqref{Gdef} of $G$. In practice, $\pi(g)$ can be easily extracted from $g$, simply by discarding the parts of $g$ which represent transformations within projective factors and only keeping the ones which permute projective factors as a whole. In summary, if $F(g)$ in Eq.~\eqref{Fdef} equals one then $g$ represents a regular symmetry transformation, otherwise it is an R-symmetry transformation.

\section{Examples}\label{sec:ex}
\vskip-6pt
In this section, we illustrate the above procedure with a number of examples, starting with a review of the quintic example in Ref.~\cite{Witten:1985xc}.

\subsection{Global Symmetries of the $\IZ_5{\times}\IZ_5$ Quintic Quotient}
\vskip-6pt
We consider the ambient space ${\cal A}=\mathbb{P}^4$ with homogenous coordinates ${\bf x}=(x^0,\ldots ,x^4)^T$ and symmetry group $G={\rm PGL}(5,\mathbb{C})$. The quintic, which is the entry with number $7890$ in the standard CICY list of Ref.~\cite{Candelas:1987kf}, is defined as the zero locus of a single degree five polynomial $p$ and is represented by the configuration matrix
\begin{equation}
X=\;\cicy{\IP^4}{5}^{1,101}_{-200}
\end{equation} 
On a five-dimensional sub-space of the $101$-dimensional complex structure moduli space the quintic has a well-known freely-acting symmetry $G_\text{f}=\mathbb{Z}_5\times\mathbb{Z}_5$. Explicitly, this symmetry can be written as $G_\text{f}=\langle S,T\rangle$ with the action of the generators $S$ and $T$ on the homogeneous coordinates ${\bf x}$ specified by the matrices
\begin{equation}
 S={\rm diag}(1,~\zeta,~\zeta^2,~\zeta^3,~\zeta^4)\;,\qquad 
 T=\left(\begin{array}{ccccc}0&0&0&0&1\\1&0&0&0&0\\0&1&0&0&0\\
                                            0&0&1&0&0\\0&0&0&1&0\end{array}\right)\; ,
\end{equation}                                            
where $\zeta$ is a fifth root of unity and the associated action on the defining polynomial $p$ trivial, so $\rho(S)=\rho(T)=1$. Note that, while these two matrices do not commute, they do commute projectively and, hence, seen as elements of ${\rm PGL}(5,\mathbb{C})$, they do indeed generate the group $\mathbb{Z}_5\times\mathbb{Z}_5$. The most general defining polynomial $p$ consistent with this symmetry is given by
\begin{align}\label{psymm}
&&p~=&~\sum_{\kappa=1}^{6} a_{\kappa}\text{J}_\kappa\text{~,~~where}\notag\\
\text{J}_1~=&~{\prod}_i x_i,&\text{J}_2~=&~{\sum}_i x_{i-1}^2~x_{i}~x_{i+1}^2,&\text{J}_3~=&~{\sum}_i x_{i-2}^2~x_{i}~x_{i+2}^2, \\[3pt]
\text{J}_4~=&~{\sum}_i x_{i-2}~x_{i}^3~x_{i+2},&\text{J}_5~=&~{\sum}_i x_{i-1}~x_{i}^3~x_{i+1},&\text{J}_6~=&~{\sum}_{i} x_i^5,\notag
\end{align}
where the $a_i$, $i=1,\ldots ,6$ are complex coefficients. We are interested in $\mathbb{Z}_5\times\mathbb{Z}_5$ symmetric quintics $X$, defined by polynomials of the above form, and their quotients $Y=X/(\mathbb{Z}_5\times\mathbb{Z}_5)$. Note that $h^{2,1}(Y)=5$ and that this five-dimensional complex structure moduli space of $Y$ is described by the projectivisation of the parameters $a_i$ in Eq.~\eqref{psymm}. We would now like to determine the generic symmetries of the quotient $Y$. 

Following our general procedure, we begin by imposing the normaliser condition~\eqref{normcond} on $5\times 5$ matrices $g\in N_G(G_\text{f})$. In terms of the generators $S$ and $T$ this condition can be stated more explicitly as
\begin{equation}
 g\,S\,g^{-1}=S^{\alpha(g)}T^{\beta(g)}\; ,\qquad g\,T\,g^{-1}=S^{\gamma(g)}T^{\delta(g)}\; , \label{quintnorm}
\end{equation}
where  $\alpha(g),\beta(g),\gamma(g),\delta(g)\in\mathbb{Z}_5$. The idea is simply that, with $g$ an element of the normaliser, conjugation of $S$ and $T$ by $g$ must lead to another $\mathbb{Z}_5\times\mathbb{Z}_5$ group element which is then parametrised by the expressions on the right-hand sides in Eq.~\eqref{quintnorm}. It is straightforward to show from Eq.~\eqref{quintnorm} that $(\alpha(g)\delta(g)-\beta(g)\gamma(g))=1\mbox{ mod }5$ and, hence, that
\begin{equation}
 M(g):=\left(\!\!\begin{array}{ll}\alpha(g)&\beta(g)\\\gamma(g)&\delta(g)\end{array}\!\!\right)\in{\rm SL}(2,\mathbb{Z}_5)\; .
\end{equation} 
In fact, the above matrices $M(g)$ provide an explicit realisation of the automorphism~\eqref{psidef} defined by each $g\in N_G(G_\text{f})$, that is, $\psi_g=M(g)$ and we have ${\rm SL}(2,\mathbb{Z}_5)\subset{\rm Aut}(\mathbb{Z}_5\times\mathbb{Z}_5)={\rm GL}(2,\mathbb{Z}_5)$. Conversely, for each matrix
\begin{equation}
 M=\left(\!\!\begin{array}{ll}\alpha&\beta\\\gamma&\delta\end{array}\!\!\right)\in {\rm SL}(2,\mathbb{Z}_5) \label{Mdef}
\end{equation}
we can solve the matrix equations
\begin{equation}
  g\,S\,g^{-1}=S^{\alpha}T^{\beta}\; ,\qquad g\,T\,g^{-1}=S^{\gamma}T^{\delta}
\label{quinticnorm}
\end{equation}
in order to find the elements $g$ of the normaliser. The centraliser $C_G(G_\text{f})$ is obtained by solving Eq.~\eqref{quinticnorm} for the matrix $M=\mathbbm{1}_2$ and found to be $C_G(G_\text{f})=G_\text{f}=\mathbb{Z}_5\times\mathbb{Z}_5$. Given an arbitrary $ {\rm SL}(2,\mathbb{Z}_5)$ matrix of the form~\eqref{Mdef}, it turns out the solutions to Eqs.~\eqref{quinticnorm} are unique up to multiplication with elements of the centraliser.  In practice, following our discussions in Section~\ref{sec:method}, we only need to solve Eqs.~\eqref{quinticnorm} for the generators of $\text{SL}(2,\IZ_5)$, which are identical to the generators of  SL$(2, \IZ)$ and are given by
 \beq
M_1=\left(\!\!\begin{array}{cc}
0 & -1 \\
1 & \+0 
\end{array}\!\!\right)\;,\quad
M_2=\left(\!\!\begin{array}{cc}
1 & 1 \\
0 & 1 
\end{array}\!\!\right)\; .
\eeq
For these two SL$(2, \IZ_5)$ matrices, the solutions $g_1$ and $g_2$ to the Eqs.~\eqref{quinticnorm}  (up to multiplication of elements in the centraliser) are
\beq
g_1=\left(
\begin{array}{ccccc}
 1 & 1 & 1 & 1 & 1 \\
 1 & \zeta  & \zeta ^2 & \zeta ^3 & \zeta ^4 \\
 1 & \zeta ^2 & \zeta ^4 & \zeta  & \zeta ^3 \\
 1 & \zeta ^3 & \zeta  & \zeta ^4 & \zeta ^2 \\
 1 & \zeta ^4 & \zeta ^3 & \zeta ^2 & \zeta  \\
\end{array}\right)\;,\quad g_2=\left(
\begin{array}{ccccc}
 1 & 1 & \zeta  & \zeta ^3 & \zeta  \\
 \zeta  & 1 & 1 & \zeta  & \zeta ^3 \\
 \zeta ^3 & \zeta  & 1 & 1 & \zeta  \\
 \zeta  & \zeta ^3 & \zeta  & 1 & 1 \\
 1 & \zeta  & \zeta ^3 & \zeta  & 1 \\
\end{array}
\right).\eeq
These matrices along with the generators $S$, $T$ of the centraliser $C_G(G_\text{f})=G_\text{f}$ generate the normaliser $N_G(G_\text{f})$, which turns out to be a group of order 3000. Imposing Eq.~\eqref{pinv}, we find only a $(\IZ_5{\times}\IZ_{5}){\rtimes}\mathbb{Z}_2$ sub-group of this group leaves all $G_\text{f}$ invariant quintics invariant and, hence, we have
\begin{equation}
 N^{\star}_G(G_\text{f})=(\IZ_5{\times}\IZ_{5}){\rtimes}\mathbb{Z}_2\; .
\end{equation}
It follows that
\begin{equation}
 G_\text{Y}=\langle g\rangle\cong \mathbb{Z}_2\; ,
\end{equation}
where the generator $g$ is explicitly given by
\begin{equation}
 g=\left(\begin{array}{lllll}0&0&0&0&1\\0&0&0&1&0\\0&0&1&0&0\\0&1&0&0&0\\1&0&0&0&0\end{array}\right)\; .
\end{equation}
This matrix $g$ solves Eq.~\eqref{quinticnorm} for $M=-\mathbbm{1}_2$.

Since we have only one projective factor, there is no permutation involved so that $\rho(g)=1$ and, since the single defining polynomial is invariant under $g$ we also have $\pi(g)=1$. Since ${\rm det}(g)=1$, it follows from Eq.~\eqref{Fdef} that
\begin{equation}
 F(g)=1\; ,
\end{equation}
so that we have found a regular $\mathbb{Z}_2$ symmetry, rather than an R-symmetry.  

\subsection{A CICY quotient with a non-Abelian symmetry}
\vskip-6pt
Our next example is for the CICY with number $14$ in the standard list of Ref.~\cite{Candelas:1987kf}, a manifold which can be viewed as a split of the bi-cubic in $\mathbb{P}^2\times\mathbb{P}^2$. It is defined in the ambient space ${\cal A}=\mathbb{P}^1\times\mathbb{P}^2\times\mathbb{P}^2$ with symmetry group
\begin{equation}
 G=S_2\ltimes\left({\rm PGL}(2,\mathbb{C})\times{\rm PGL}(3,\mathbb{C})\times {\rm PGL}(3,\mathbb{C})\right)\; ,
\end{equation}
where the $S_2$ group permutes the two $\mathbb{P}^2$ factors of the ambient space. The homogeneous coordinates of ${\cal A}$ are denoted by ${\bf x}=(x_0,x_1,y_0,y_1,y_2,z_0,z_1,z_2)^T$. The CICY is defined as the common zero locus of two polynomials, $p_1$ and $p_2$, whose degrees are specified by the configuration matrix
\beq
X\;=\;\cicy{\IP^1\\ \IP^2\\ \IP^2}{1&1\\ 3&0\\ 0&3}^{19,19}_{0}\; .
\label{SplitBicubic1}\eeq
On a three-dimensional sub-space of the $19$-dimensional complex structure moduli space, this manifold has a freely-acting symmetry $G_\text{f}=\mathbb{Z}_3{\times}\mathbb{Z}_3=\langle S,T\rangle$, whose generators $S$, $T$ act on the homogeneous coordinates ${\bf x}$ as
\begin{equation}
 S=\left(\begin{array}{lll}\mathbbm{1}_2&0&0\\0&D(\omega)&0\\0&0&D(\omega)\end{array}\right)\;,\quad
 T=\left(\begin{array}{lll}\mathbbm{1}_2&0&0\\0&P&0\\0&0&P\end{array}\right)\;,\quad
 P=\left(\begin{array}{lll}0&0&1\\1&0&0\\0&1&0\end{array}\right)\; ,
\end{equation} 
where $D(\omega)={\rm diag}(1,\omega^2,\omega)$ and $\omega$ is a nontrivial cube root of unity. These actions on the coordinates are combined with trivial actions on the polynomials, so that $\rho(S)=\rho(T)=\mathbbm{1}_2$. The most general defining polynomials consistent with this symmetry are
\begin{equation}
\begin{array}{lll}
p_1 &=& (a_1 x_0+a_2 x_1)~z_0 z_1 z_2+(a_3 x_0+a_4 x_1)~( z_0^3+ z_1^3+ z_2^3)\\[4pt]
p_2 &=& (a_5 x_0+a_6 x_1)~y_0 y_1 y_2+(a_7 x_0+a_8 x_1)~( y_0^3+ y_1^3+ y_2^3)\; ,
\end{array}
\end{equation}
where $a_i\in\mathbb{C}$ are arbitrary coefficients. We are interested in $\mathbb{Z}_3{\times}\mathbb{Z}_3$ symmetric manifolds $X$ defined by polynomials of this type and the associated quotients $Y\!\!=\!\!X/\mathbb{Z}_3{\times}\mathbb{Z}_3$, whose Hodge number is $h^{2,1}(Y)=3$. 

We begin by computing the centraliser $C^*_G(G_\text{f})$ by solving the normaliser condition~\eqref{auteqn} for $\psi=\text{id}$ and then imposing the invariance condition~\eqref{pinv}. This yields
\beq
C^*_G(G_\text{f})=~\IZ_3^4=\langle S,~T,~ g_1,~g_2\rangle~,
\eeq
where the generators $g_i$ act on the homogeneous co-ordinates of ${\cal A}$ as,
\begin{align}
g_1=\left(\begin{array}{ccc}\mathbbm{1}_2&0&0\\0&P&0\\0&0&P\,D(\omega^2)\end{array}\right)\;,\quad
 g_2=\left(\begin{array}{lll}\mathbbm{1}_2&0&0\\0&P&0\\0&0&\mathbbm{1}_3\end{array}\right)\; .
\end{align}
To compute the normaliser, we first note that $\text{Aut}(G_\text{f})=\text{GL}(2,\IZ_3)$, a group of order 48. Unlike in the case of the previous example of the quintic quotient, we do not know at this stage, automorphisms from which subgroup of $\text{Aut}(G_\text{f})$ will yield solutions to the normaliser condition~\eqref{auteqn}. Therefore, we solve Eq.~\eqref{auteqn} using all automorphisms $\psi$ from a minimal list of generators of all subgroups of $\text{GL}(2,\IZ_3)$, and then impose the invariance condition~\eqref{pinv}. These solutions are then combined with the elements of the restricted centraliser $C^*_G(G_\text{f})$ to generate the normaliser $N^*_G(G_\text{f})$. We arrive at,
\beq
N^*_G(G_\text{f})=~\IZ_3^4{\rtimes}\IZ_2=\langle S,~T,~g_1,~g_2,~g_3\rangle~,
\eeq
where the generator $g_3$ acts on ${\cal A}$ as,
\begin{align}
g_3=\left(\begin{array}{lll}\mathbbm{1}_2&0&0\\0&\widetilde{P}&0\\0&0&\omega\widetilde{P}\end{array}\right)\; 
\quad\text{with}\quad \widetilde{P}=\left(\begin{array}{lll}0&0&1\\0&\omega^2&0\\\omega&0&0\end{array}\right)\; .
\end{align}
This is a non-Abelian group of order $162$. In order to find the symmetry group of $Y$ we take the quotient by $G_\text{f}=\mathbb{Z}_3{\times}\mathbb{Z}_3$ to find
\begin{equation}
 G_\text{Y}=\IZ_3 ^ 2 {\rtimes} \IZ_2=\langle g_1,~g_2,~g_3\rangle\; ,
\end{equation}
a non-Abelian group of order $18$. Each generator is combined with the trivial action on the polynomials, so $\rho(g_i)=\mathbbm{1}_2$, for $i=1,2,3$. Since the above generators do not permute any projective factors we have $\pi(g_i)=\mathbbm{1}_3$ and since ${\rm det}(g_i)=1$ we have from Eq.~\eqref{Fdef} that
\begin{equation}
 F(g_i)=1
\end{equation}
for $i=1,2,3$. This means that the symmetry group $G_\text{Y}=\IZ_3 ^ 2 {\rtimes} \IZ_2$ is a regular symmetry, rather than an R-symmetry.  

\subsection{A CICY quotient with an R-symmetry}
\vskip-6pt
This example is for a co-dimension four CICY, number $7861$ in the list of Ref.~\cite{Candelas:1987kf}, defined in the ambient space ${\cal A}=\mathbb{P}^7$  with coordinates ${\bf x}=(x_0,\ldots ,x_7)^T$ and symmetry group
\begin{equation}
 G={\rm PGL}(8,\mathbb{C})\; .
\end{equation}
The CICY $X\subset{\cal A}$ is given by the common zero locus of four quadrics $p_i$, where $i=1,2,3,4$, and it is characterised by the configuration matrix 
\begin{equation}
 X\;=\;\cicy{\IP^7}{2&2&2&2}^{1,65}_{-128}\; .
\end{equation} 
On a nine-dimensional subspace of the $65$-dimensional complex structure moduli space, this CICY has a freely-acting $G_\text{f}=\mathbb{Z}_2\times \mathbb{Z}_2\times \mathbb{Z}_2=\langle S,T,U\rangle$ symmetry with the action of the three generators on the homogeneous coordinates ${\bf x}$ given by
\begin{equation}
\begin{array}{lll}
 S&=&{\rm diag}(1,-1,\+1,-1,\+1,-1,\+1,-1)\\
 T&=&{\rm diag}(1,-1,-1,\+1,\+1,-1,-1,\+1)\\
 U&=&{\rm diag}(1,-1,-1,-1,-1,\+1,\+1,\+1)\; ,
\end{array} 
\end{equation} 
and corresponding trivial actions on the defining polynomials, that is, $\rho(S)=\rho(T)=\rho(U)=\mathbbm{1}_4$. The most general set of defining equations consistent with this symmetry is given by
\begin{equation}
p_i~=~\sum_{j} a_{i j} x_j^2\;,
\end{equation}
for  $i \in \{1,2,3,4\}$, where $a_{ij}\in\mathbb{C}$ are arbitrary coefficients. We would like to consider $G_\text{f}=\mathbb{Z}_2^3$ invariant CICY manifolds $X$ defined by polynomials of this form and the associated quotients $Y=X/\mathbb{Z}_2^3$. 

In this example, $\text{Aut}(G_\text{f})=\text{PSL}(3,\IZ_2)$, a group of order 168. Imposing the normaliser condition~\eqref{auteqn} using $\psi\in {\text{PSL}(3,\IZ_2)}$ and following the prescription in Section~\ref{sec:method}, as well as the invariance~\eqref{pinv} of the defining equations, we find 
\begin{equation}
 C^\star_G(G_\text{f})=N^\star_G(G_\text{f})=\mathbb{Z}_2^7=\langle S,~T,~U,~g_1,~g_2,~g_3,~g_4\rangle
\end{equation}
where the four generators $g_i$ of this symmetry act on the homogeneous coordinates as,
\begin{equation}
\begin{array}{lll}
 g_1&=&{\rm diag}(1,-1,-1,-1,-1,\+1,\+1,-1)\\
 g_2&=&{\rm diag}(1,-1,-1,-1,-1,\+1,-1,\+1)\\
 g_3&=&{\rm diag}(1,-1,-1,-1,-1,-1,\+1,\+1)\\
 g_4&=&{\rm diag}(1,-1,-1,-1,\+1,\+1,\+1,\+1)\; .
\end{array}
\end{equation} 
For the symmetry group of $Y$ we divide by $G_\text{f}=\mathbb{Z}_2^3$ and find
\begin{equation}
 G_\text{Y}=\mathbb{Z}_2^4=\langle g_1,~g_2,~g_3,~g_4\rangle\; .
\end{equation} 
The $g_i$ all have trivial actions on the polynomials, so $\rho(g_i)=\mathbbm{1}_4$. There is no permutation of projective factors involved, so that $\pi(g_i)=1$ and, since ${\rm det}(g_i)=-1$ it follows from Eq.~\eqref{Fdef} that
\begin{equation}
 F(g_i)=-1\; ,
\end{equation}
for $i=1,2,3,4$. This means that $G_\text{Y}=\mathbb{Z}_2^4$ is, in fact, an R-symmetry.

\section{Practical implementation and results} \label{sec:results}
\vskip-6pt
We would like to find the symmetry groups, $G_\text{Y}$, for all $1695$ CICIY quotients which can be constructed from the standard CICY list in Ref.~\cite{Candelas:1987kf} and the classification of freely-acting symmetries in Ref.~\cite{Braun:2010vc}. The data for both the CICY manifolds and the freely-acting symmetries is available for download and we will use the version of this dataset available at \cite{cicylist}. An entry in this dataset consists of a pair, $(X,G_\text{f})$, of a CICY manifold and a freely-acting symmetry. Following the methods explained in Section~\eqref{sec:method}, we will then compute, for each such pair $(X,G_\text{f})$, the groups ${\rm Aut}(G_\text{f})$, $C^\star_G(G_\text{f})$, $N^\star_G(G_\text{f})$ and the symmetry group $G_\text{Y}$ of the CICY quotient $Y=X/G_\text{f}$. 

The practical implementation of this computation involves the dataset~\cite{cicylist}, the package GAP~\cite{gap} for many of the group-theoretical tasks, such as computing the automorphism groups ${\rm Aut}(G_\text{f})$, and Mathematica together with the CICY package~\cite{cicypackage} for all remaining tasks. 

The results can be summarised as follows. Of the 1695 CICY quotients $Y$, a total of 381 were found to admit nontrivial generic discrete symmetry groups $G_\text{Y}$. Of these, 113 CICY quotients have an R-symmetry (which, in some cases, consists only of a $\mathbb{Z}_2$ sub-group of the full symmetry group $G_\text{Y}$) and 187 CICY quotients have a symmetry group $G_\text{Y}=\IZ_2$. Eight further groups $G_\text{Y}$, with a maximal group order of $18$, appear within the dataset, and the full list of possibilities is
\begin{equation}
 G_\text{Y}\in\left\{ \IZ_2,~\IZ_3,~\IZ_4,~\IZ_2^2,~\IZ_2^3,~\ID_8,~\IZ_2^4,~\IZ_2{\times}\ID_8,~(\IZ_3{\times}\IZ_3){\rtimes}\IZ_2\right\}\; .
\end{equation} 
The frequency with which each of these groups appears in the dataset is provided in Table~\ref{tab:NewGroupLis}.
\begin{table}[H]
\begin{center}
\vspace{5pt}
\begin{tabular}{ |c ||>{\hskip4pt} c<{\hskip4pt}| >{\hskip4pt} c<{\hskip4pt} | >{\hskip4pt} c<{\hskip4pt} |>{\hskip4pt} c<{\hskip4pt} | >{\hskip4pt} c<{\hskip4pt} | >{\hskip4pt} c<{\hskip4pt}|>{\hskip4pt} c<{\hskip4pt}|c|c| }
\hline
\varstr{17pt}{14pt} $G_\text{Y}$ &
\varstr{17pt}{14pt} {$\IZ_2$} &
\varstr{17pt}{14pt} {$\IZ_3$} &
\varstr{17pt}{14pt} {$\IZ_4$}  &
\varstr{17pt}{14pt} {$\IZ_2^2$}  &
\varstr{17pt}{14pt} {$\IZ_2^3$}  &
\varstr{17pt}{14pt} {$\ID_8$}  &
\varstr{17pt}{14pt} {$\IZ_2^4$} &
\varstr{17pt}{14pt} {$\IZ_2{\times}\ID_8$}  &
\varstr{17pt}{14pt} {$(\IZ_3{\times}\IZ_3){\rtimes}\IZ_2$}  
\\ \hline\hline
\varstr{14pt}{8pt} \text{~Regular symmetries~} & 155 & 35 & 5 & 31 & 36 & 11 & 0 & 2 & 4 \\\hline
\varstr{14pt}{8pt} \text{~R-symmetries~} & 52 & 0 & 0 & 33 & 25 & 0 & 3 & 0 & 0  \\\hline
\end{tabular}
\caption{\it Symmetry groups of CICY quotients and their frequency.}\label{tab:NewGroupLis}
  \end{center}
   \vskip -0.7cm
 \end{table}
A more detailed account of the results for all $381$ non-trivial cases can be found in Table~\ref{tab:SymmetryGroups} in Appendix~\ref{app:results}. The table lists the freely-acting symmetry group $G_\text{f}$ in the first column, provides a list of identifiers (CICY \#,~SYMM \#) for pairs $(X,G_\text{f})$ which indicate the position in the dataset~\cite{cicylist} in the second column and provides the groups $C_G^\star(G_\text{f})$, $N_G^\star(G_\text{f})$ and $G_\text{Y}=N^\star_G(G_\text{f})/G_\text{f}$ in the remaining three columns. The data for the matrix generators $g$ acting on the homogeneous ambient space coordinates and the matrices $\rho(g)$ for the corresponding actions on the polynomials for all these symmetries is too lengthy to be reproduced on paper but it can be downloaded from Ref.~\cite{data}. A table with the automorphism groups ${\rm Aut}(G_\text{f})$ of the freely-acting symmetry groups $G_\text{f}$ which enter the computation can be found in the appendix. 

\section{Conclusion}\label{sec:conclusion}
\vskip-6pt
In this paper, we have considered CICY quotients $Y=X/G_\text{f}$, obtained as quotients of CICY manifolds $X$ by freely-acting symmetries $G_\text{f}$, and we have studied the symmetries $G_\text{Y}$, freely and non-freely acting, of these quotients. Such symmetries $G_\text{Y}$ can lead to discrete gauge symmetries in low-energy theories obtained by string compactification on $Y$ and are, therefore, of phenomenological relevance. More specifically, we have focused on those $G_\text{Y}$ which are symmetries everywhere in the moduli space of the quotient $Y$. Only those ``generic" symmetries can lead to low-energy symmetries which are manifest for all values of the moduli fields, rather than just for special values of those fields. 

The experience so far, particularly from the classification of freely-acting symmetries, suggests that such generic symmetries of CY manifolds typically do not exist. Put simply, CY manifolds are too complicated to display symmetries at a generic point in moduli space - only at lower-dimensional sub-loci in moduli space do symmetries appear. However, this expectation is derived from the study of (freely-acting) symmetries for CY manifolds with a non-trivial first fundamental group. The main result of the present paper is that the situation is quite different for CY quotient manifolds and non-freely acting symmetries. Our classification strongly suggests that generic, non-freely acting symmetries for CY quotients arise relatively frequently. For the $1695$ CICY quotients $Y=X/G_\text{f}$ which can be constructed from the CICY manifolds $X$ in the standard list~\cite{Candelas:1987kf} and freely-acting symmetries $G_\text{f}$ as classified in Ref.~\cite{Braun:2010vc} we find such generic, non-freely acting symmetries on about $23\%$ of these quotient manifolds. This figure should, for example, be compared with the frequency of freely-acting symmetries for CICY manifolds which, from the classification of Ref.~\cite{Braun:2010vc}, stands at about $2.5\%$, but with each of these symmetries appearing only at non-generic points in moduli space. 

CY quotient manifolds are the preferred compactification manifolds for realistic model building in the context of the heterotic string. Hence, our results suggest that low-energy discrete symmetries which originate from the compactification space are a common occurrence for heterotic string models. 

On the $381$ CICY quotients $Y$ with non-trivial generic symmetry group, we find that $9$ different symmetry groups $G_\text{Y}$ can arise, namely
\[
 G_\text{Y}\in\left\{ \IZ_2,~\IZ_3,~\IZ_4,~\IZ_2^2,~\IZ_2^3,~\ID_8,~\IZ_2^4,~\IZ_2{\times}\ID_8,~(\IZ_3{\times}\IZ_3){\rtimes}\IZ_2\right\}\; .
\]
For $113$ of those CICY quotients all or part of $G_\text{Y}$ corresponds to an R-symmetry, for the others $G_\text{Y}$ is a regular symmetry. 

There are several obvious extensions of the present work. In the present paper, we have classified symmetries $G_\text{Y}$ which leave the CICY quotients $Y$ invariant for each choice of complex structure. This means that resulting low-energy discrete symmetries will act trivially on the complex structure moduli. It is also possible to consider symmetries which map between manifolds $Y$ corresponding to different choices of complex structure, leading to low-energy symmetries with a non-trivial action on the complex structure moduli. We expect that such symmetries can be found by methods similar to the ones described here, subject to a suitable modification of the invariance condition~\eqref{pinv} for the defining polynomials. Another possible extension would be to find non-generic symmetries which only arise at a sub-locus in the complex structure moduli space of a CICY quotient. Since the present method heavily relies on the invariance of the {\em entire} family of polynomials describing the quotient CICY, finding such non-generic symmetries will likely require a different set of methods, possibly a modification of the approach taken in Ref.~\cite{Braun:2010vc}. For the specific case of the quintic CY, work in this direction is under way~\cite{wip}. It would also be interesting to know if results similar to the present ones arise for free quotients of CY manifolds constructed as hyper-surfaces in toric four-folds. However, this requires a classification of freely-acting symmetries for these CY manifolds which, to date, has been achieved only partially~\cite{Batyrev:2005jc,Braun:2017juz}.

Finally, the symmetries found in this paper may be of direct relevance for the heterotic line bundle standard models on CICY quotients found in Ref.~\cite{Anderson:2011ns,Anderson:2012yf}. It would be interesting to analyse this in more detail and, in particular, check if some of the present symmetries lift to the gauge bundle.\\

{\bf Acknowledgements} The authors would like to acknowledge Philip Candelas for several helpful discussions. A.L. is partially supported by the EPSRC network grant EP/N007158/1 and by the STFC grant~ST/L000474/1. 
\newpage


\newpage
\begin{appendix}
\section{Automorphism groups}\label{Appendix:AutGrs}
\vskip-6pt
The automorphism group ${\rm Aut}(G)$ of a group $G$ is the set of all group automorphisms $\psi:G\rightarrow G$ which forms a group under the composition of maps. In the main part of the paper, we consider quotients $Y=X/G_\text{f}$ of CICY manifolds $X$ by groups $G_\text{f}$ which act freely on $X$. The computation of the symmetry groups $G_\text{Y}$ of these CICY quotients requires the automorphism groups ${\rm Aut}(G_\text{f})$ for all freely-acting groups $G_\text{f}$ which arise in the classification of Ref.~\cite{Braun:2010vc}. These automorphism groups can be computed with the package GAP~\cite{gap} and the results for all relevant groups $G_\text{f}$ are listed in Table~\ref{tab:AutGroupsNA} below.
\begin{center}
\begin{longtable}{|c | c c | c c|}
\captionsetup{width=0.96\textwidth}
\caption{\label{tab:AutGroupsNA}{\it Automorphism groups ${\rm Aut}(G_\text{f})$ of groups $G_\text{f}$ acting freely on CICY manifolds, computed using GAP. For convenience, we also list the GAP identifiers for all groups, a pair of two numbers, the first of which represents the group order. For some groups with large order, the complete identifier or the structure description of the automorphism group was not available.}}\\
\hline 
\str\textbf{~~\#~~} & \textbf{$\text{G}_\text{f}$}  & \textbf{GAP ID}~ & \textbf{\hfill Aut($\text{G}_\text{f}$)\hfill} & \textbf{GAP ID} \\ \hline \hline
\endfirsthead
\hline 
\str\textbf{~~\#~~}&\textbf{$\text{G}_\text{f}$}  & \textbf{GAP ID}~ & \textbf{\hfill Aut($\text{G}_\text{f}$)\hfill} & \textbf{GAP ID} \\ \hline \hline
\endhead
%
\hline \multicolumn{5}{|r|}{{\str Continued on next page}} \\ \hline

\endfoot
%
\multicolumn{5}{|c|}{\str}\\ \hline
\endlastfoot
 \varstr{16pt}{12pt} 1& \hspace{0.6cm} $\IZ_2$ \hspace{0.6cm} & [2, 1] & $\mathbbm{1}$ & [1, 1] \\ 					
\hline \varstr{16pt}{12pt} 2&$\IZ_3$ & [3, 1] & $\IZ_2$ & [2, 1] \\ 				
\hline \varstr{16pt}{12pt} 3&$\IZ_4$ & [4, 1] & $\IZ_2$ & [2, 1]  \\ 				
\hline \varstr{16pt}{12pt} 4&$\IZ_2{\times}\IZ_2$ & [4, 2] & $\IS_3$ & [6, 1] \\			
\hline \varstr{16pt}{12pt} 5&$\IZ_5$ & [5, 1] & $\IZ_4$ & [4, 1] \\
\hline \varstr{16pt}{12pt} 6&$\IZ_6$ & [6, 2] & $\IZ_2$ & [2, 1] \\
\hline \varstr{16pt}{12pt} 7&$\IZ_8$ & [8, 1] & $\IZ_2{\times}\IZ_2$ & [4, 2] \\		
\hline \varstr{16pt}{12pt} 8&$\IZ_4{\times}\IZ_2$ & [8, 2] & $\ID_8$ & [8, 3] \\
\hline \varstr{16pt}{12pt} 9&$\IZ_2^3$ & [8, 5] & PSL(3,2) & [168, 42] \\		
\hline \varstr{16pt}{12pt} 10&$\IQ_8$ & [8,4] & $\text{S}_4$ & [24, 12]\\
\hline \varstr{16pt}{12pt} 11&$\IZ_3{\times}\IZ_3$ & [9, 2] & GL(2,3) & [48, 29] \\
\hline \varstr{16pt}{12pt} 12&$\IZ_{10}$ & [10, 2] & $\IZ_4$ & [4, 1] \\
\hline \varstr{16pt}{12pt} 13&$\text{Dic}_3$ & [12, 1] & $\ID_{12}$ & [12, 4]\\
\hline \varstr{16pt}{12pt} 14&$\IZ_{12}$ & [12, 2]  & $\IZ_2{\times}\IZ_2$ & [4, 2] \\ 
\hline \varstr{16pt}{12pt} 15&$\IZ_4{\times}\IZ_4$ & [16, 2] & $(\IZ_2^2{\times}\IA_4){\rtimes}\IZ_2$ & [96, 195] \\
\hline \varstr{16pt}{12pt} 16&$\IZ_8{\times}\IZ_2$ & [16, 5] & $\IZ_2{\times}\ID_8$ & [16, 11] \\
\hline \varstr{16pt}{12pt} 17&$\IZ_4{\times}\IZ_2^2$ & [16, 10]  & $
[ ((\IZ_2 {\times} \ID_8) {\rtimes}\IZ_2){\rtimes}\IZ_3 ] {\rtimes}\IZ_2$ & [192, 1493]\\
\hline \varstr{16pt}{12pt} 18&$\IZ_4{\rtimes}\IZ_4$ & [16, 4] & $\IZ_2^4{\rtimes} \IZ_2$ & [32, 27]\\
\hline \varstr{16pt}{12pt} 19 & $\IZ_8{\rtimes}\IZ_2$ & [16, 6] & $\IZ_2{\times} \ID_8$ & [16, 11] \\
\hline \varstr{16pt}{12pt} 20 & $\IZ_2{\times}\IQ_8$ & [16, 12] & $((\IZ_2^4{\rtimes} \IZ_3){\rtimes} \IZ_2){\rtimes} \IZ_2$ & [192, 955] \\
\hline \varstr{16pt}{12pt} 21 & $\IZ_{10}{\times}\IZ_2$ & [20, 5] & $\IZ_4{\times}\IS_3$ & [24, 5] \\
\hline \varstr{16pt}{12pt} 22 & $\IZ_5{\times}\IZ_5$ & [25, 2] & GL(2,5) & [480, 218] \\
\hline \varstr{16pt}{12pt} 23 & $(\IZ_4{\times}\IZ_2){\rtimes}\IZ_4$ & [32, 2] & $(\IZ_2{\times} \IZ_2 {\times} (\IZ_2 ^4 {\rtimes} \IZ_3)){\rtimes} \IZ_2$ & [384, 20100]\\
\hline \varstr{16pt}{12pt} 24 & $\IZ_8{\rtimes}\IZ_4$ & [32, 4] & 
$[\IZ_2 {\times} (((\IZ_4 {\times} \IZ_2) {\rtimes} \IZ_2) $
  & [128, 753] \\
\hline \varstr{16pt}{12pt} 25 & $(\IZ_8{\times}\IZ_2){\rtimes}\IZ_2$ & [32, 5] & $\IZ_2 {\times} (\IZ_2^4 {\rtimes} \IZ_2)$ & [64, 202] \\
\hline \varstr{16pt}{12pt} 26 & $\IZ_8{\rtimes}\IZ_4$ & [32, 13] & $(\IZ_2^3 {\times} \ID_8) {\rtimes} \IZ_2$ & [128, 1735] \\
\hline \varstr{16pt}{12pt} 27 & $\IZ_2{\times}(\IZ_4{\rtimes}\IZ_4)$ & [32, 23] &  $[(( \IZ_2 {\times} \IZ_2 {\times} (( \IZ_4 {\times} \IZ_2)  {\rtimes} \IZ_2)) {\rtimes} \IZ_2 ) {\rtimes} \IZ_2]{\rtimes} \IZ_2$ & [512, *] \\
\hline \varstr{16pt}{12pt} 28 & $\IZ_4{\rtimes}\IQ_8$ & [32, 35] & $[( \IZ_2 {\times} ((((\IZ_4 {\times} \IZ_2) {\rtimes} \IZ_2)  
{\rtimes} \IZ_2){\rtimes} \IZ_2)) {\rtimes} \IZ_2]{\rtimes} \IZ_2$ & [512, *] \\
\hline \varstr{16pt}{12pt} 29 & $\IZ_2{\times}\IZ_2{\times}\IQ_8$ & [32, 47] & * & [9216, *] \\
\hline \varstr{16pt}{12pt} 30 & $\IZ_8{\times}\IZ_4$ & [32, 3] & $[\IZ_2 {\times} (((\IZ_4 {\times} \IZ_2) {\rtimes} \IZ_2) 
 {\rtimes} \IZ_2)] {\rtimes} \IZ_2$ & [128, 753] \\
\hline \varstr{16pt}{12pt} 31 & $\IZ_4^2{\times}\IZ_2$ & [32, 21] &  $[((\IZ_2^2 {\times} (\IZ_2^4 {\rtimes} \IZ_2)) 
{\rtimes} \IZ_2) {\rtimes} \IZ_3] {\rtimes} \IZ_2$ & [1536, *] \\
\hline\hline
\end{longtable}
\end{center}

\newpage
\section{Symmetries of CICY quotients}\label{app:results}
\begin{center}
\begin{longtable}{|>{\hskip-5pt}c<{\hskip-5pt}|c|c|c|c|>{\hskip-5pt}c<{\hskip-5pt}|}
\captionsetup{width=0.96\textwidth}
\caption{\label{tab:SymmetryGroups}{\it Global symmetry groups of smooth CICY quotients. A pair $(X,G_\text{f})$ of a CICY $X$ and a freely-acting symmetry $G_\text{f}$ is referred to by the numbers (CICY \#,~SYMM \#). Further, the centralizer $C^{*}_G(G_\text{f})$, the normalizer $N^{*}_G(G_\text{f})$ and the generic symmetry group $G_\text{Y}=N^{*}_G(G_\text{f})/G_\text{f}$ of the quotient $Y=X/G_\text{f}$ is listed. Bold numbers {(CICY \#,~SYMM \#)} indicate manifolds with global R-symmetries. For manifolds that appear with a superscript $\dagger$, only a $\IZ_2$ subgroup of the entire global symmetry group $N^{*}_G(G_\text{f})/G_\text{f}$ is an R-symmetry.}}\\
\hline 
\str\textbf{$\text{G}_\text{f}$} & \str\textbf{\textbf{{(CICY\#, SYMM\#)}}}  & {$\textbf{C}^{*}_\textbf{G} (\textbf{G}_\text{f})$} & {$\textbf{N}^{*}_\textbf{G} (\textbf{G}_\text{f})$} & {$\textbf{N}^{*}_\textbf{G} (\textbf{G}_\text{f})/\textbf{G}_\textbf{f}$}   \\ \hline \hline
\endfirsthead
\hline
\str\textbf{$\text{G}_\text{f}$} & \str\textbf{\textbf{{(CICY\#, SYMM\#)}}}  & {$\textbf{C}^{*}_\textbf{G} (\textbf{G}_\text{f})$} & {$\textbf{N}^{*}_\textbf{G} (\textbf{G}_\text{f})$} & {$\textbf{N}^{*}_\textbf{G} (\textbf{G}_\text{f})/\textbf{G}_\textbf{f}$}   \\ \hline\hline
\endhead
\hline\hline \multicolumn{5}{|r|}{{\str Continued on next page}} \\ \hline
\endfoot
\hline\hline\multicolumn{5}{|c|}{\str}\\ \hline
\endlastfoot
 \multirow{2}{*}{\raisebox{-0.5cm}{$\IZ_2$}} & \begin{minipage}[c][50pt][c]{1.3in}
\begin{gather*}
(19, 1), (21, 3), (27, 1), \\
(28, 2), (30, 1) \\ 
\end{gather*}
\end{minipage} & $\IZ_2 ^2$ & $\IZ_2 ^2$ & $\IZ_2$ \\
\varstr{17pt}{14pt} & $\mathbf{({{6836, {11}}})}$ & $\IZ_2 ^ 4$ & $\IZ_2 ^ 4$ & $\IZ_2 ^3$  \\ \hline
\multirow{2}{*}{\raisebox{-0.5cm}{$\IZ_3$}} & \varstr{17pt}{14pt} $(6, 33)$  & $\IZ_3$ & $\IS_3$ & $\IZ_2$ \\
& \varstr{17pt}{14pt} ${{(14, {1})}, {(18, {1})}, {(26, {1})}}$ & $\IZ_3 ^2$ & $\IZ_3 ^2$ & $\IZ_3$ \\ \hline
\multirow{3}{*}{\raisebox{-2.8cm}{$\IZ_4$}} &\varstr{17pt}{14pt} $({{19, {4}}), ({20, {5}}})$ & $\IZ_4 ^2$ & $\IZ_4 ^2$ & $\IZ_4$  \\ 
& \begin{minipage}[c][50pt][c]{1.3in}
\begin{gather*}
(19, 7- 8), (20, 3-6),\\
(21, 7- 8) \\ 
\end{gather*}
\end{minipage} & $\IZ_4 {\times} \IZ_2$ & $\IZ_4 {\times} \IZ_2$ & $\IZ_2$ \\
& \varstr{17pt}{14pt} \begin{minipage}[c][50pt][c]{1.0in}
\begin{gather*}
(19, 9), (21, 9)\\
(30, 4), (2568, 8)\\ 
\end{gather*}
\end{minipage} & $\IZ_4$ & $\ID_8$ & $\IZ_2$ \\
& \varstr{17pt}{14pt} $({{21, {6}}})$ & $\IZ_4 ^2$ & $\IZ_4 ^2 {\rtimes} \IZ_2$ & $\ID_8$\\ 
& \varstr{17pt}{14pt} $({{6836, {14}}})$ & $\IZ_4 {\times} \IZ_2$ & $\IZ_2 {\times} \ID_8$ & $\IZ_2 ^2$  \\ \hline
\multirow{5}{*}{\raisebox{-1.5cm}{$\IZ_2{\times}\IZ_2$}} & \varstr{17pt}{14pt} $\mathbf{({{19, {10}})}, \mathbf{({20, {8}})}, \mathbf{({21, {10}}})}$\footnote{For this manifold, only a $\IZ_2{\times}\IZ_2$ subgroup of the entire $N^{*}_G(G_\text{f})/G_\text{f}=\IZ_2^4$ is an R-symmetry.} & $\IZ_2 ^ 6$ & $\IZ_2 ^ 6$ & $\IZ_2 ^ 4$ \\ 
&\varstr{17pt}{14pt} \begin{minipage}[c][90pt][c]{1.5in}
\begin{gather*}
\mathbf{(19, 11-16)}, \\
\hskip-6pt(\mathbf{6836}, \{ \mathbf{15-17, 30-31,} \\
\hskip-6pt\mathbf{38, 42, 46, 50, 52, 59, 62,}\\
\mathbf{69, 71, 74, 85, 88, 92}\})  \\ 
\end{gather*}
\end{minipage} & $\IZ_2 ^ 4 {\times} \IZ_2$ & $\IZ_2 ^ 4 {\times} \IZ_2$ & $\IZ_2 ^3$\\
&\varstr{17pt}{14pt} \begin{minipage}[c][120pt][c]{1.5in}
\begin{gather*}
\hskip-7pt\mathbf{(20, 9-14)}, (\mathbf{6836}, \{\mathbf{18^{\dagger}}-\\
\hskip-7pt\mathbf{20, 22, 24, 27, 32^{\dagger},33,34,}\\
\hskip-7pt\mathbf{36, 39^{\dagger},40^{\dagger},41^{\dagger}, 43, 44^{\dagger},} \\
\hskip-8pt\mathbf{45, 47, 49^{\dagger}, 51, 54, 55, 57^{\dagger},} \\
\hskip-7pt\mathbf{58, 61, 63, 66, 68^{\dagger}, 70, 73,} \\
\hskip-8pt\mathbf{77, 78, 81, 83^{\dagger}, 84, 87, 91\!}\}) \\ 
\end{gather*}
\end{minipage} & $\IZ_2 ^ 4$ & $\IZ_2 ^ 4$ & $\IZ_2 ^2$\\ 
&\varstr{17pt}{14pt} \begin{minipage}[c][200pt][c]{1.55in}
\begin{gather*}
(21, 11-16), (2564, 4),\\
\mathbf{(2566, 4-10)}, (2568, \{9\\
-36, 40\}),\!\mathbf{(5302, 5-20)}, \\
\mathbf{(6788, 4-6)}, (6836, \{21,\\
23, 25, 26, 28, 29, 35, 37,\\
48, 53, 56, 60, 64, 65, 67,\\
72, 75, 76, 79, 80, 82, 86,\\
89, 90\}), \mathbf{(7491, 5-19)},\\
(7735, \{4, 5\}), \mathbf{(7823, 2)},\\
(7861, 3) \\ 
\end{gather*}
\end{minipage} & $\IZ_2 ^3$ & $\IZ_2 ^3$ & $\IZ_2$  \\ \hline
\multirow{1}{*}{$\IZ_6$} & \varstr{17pt}{14pt}  $(6, 34-41)$ & $\IZ_6$ & $\ID_{12}$ & $\IZ_2$ \\ \hline
\varstr{17pt}{14pt} \multirow{2}{*}{\raisebox{-0.5cm}{$\IZ_8$}} & $({{19, {17}}), ({6836, {93}}})$ & $\IZ_8$ & $\IZ_8 {\rtimes} \IZ_2$ & $\IZ_2$ \\ 
& \varstr{17pt}{14pt} $({{21, {17}}})$ & $\IZ_8$ & $(\IZ_2 {\times} \ID_8) {\rtimes} \IZ_2$ & $\IZ_2 ^2$  \\ \hline
\varstr{17pt}{14pt} \multirow{2}{*}{\raisebox{-2.8cm}{$\IZ_4{\times}\IZ_2$}} & \varstr{17pt}{14pt} \begin{minipage}[c][70pt][c]{1.5in}
\begin{gather*}
(19, \{18,20\}), (2564, 6),\\
(6836, \{95, 97, 101, 103, \\
109, 111\})\\
\end{gather*}
\end{minipage} & $\IZ_4 {\times} \IZ_2 ^2$ & $\IZ_2 ^2 {\times} \ID_8$ & $\IZ_2 ^2$ \\
& \varstr{17pt}{14pt} $({{19, {19}}), ({21, {31}}})$ & $\IZ_4 {\times} \IZ_2 ^2$ & $\IZ_2 {\times} (\IZ_2 ^ 4 {\rtimes} \IZ_2)$ & $\IZ_2 ^3$ \\
& \varstr{17pt}{14pt} \begin{minipage}[c][50pt][c]{1.5in}
\begin{gather*}
(21,\{18-20, 26\}),\\
(7861, 5) \\ 
\end{gather*}
\end{minipage} & $\IZ_4 {\times} \IZ_2 ^2$ & $\IZ_4 {\times} \IZ_2 ^2$ & $\IZ_2$\\
& \varstr{17pt}{14pt} $({{21, {21}}})$ & $\IZ_4 ^2 {\times} \IZ_2 ^2$ & $\IZ_2 ^2 {\times} (\IZ_4 ^2 {\rtimes} \IZ_2)$ & $\IZ_2 {\times} \ID_8$  \\ 
& \varstr{17pt}{14pt} $({{21, \{22, 24\}}})$ & $\IZ_4 {\times} \IZ_2 ^3$ & $\IZ_2 ^3 {\times} \ID_8$ & $\IZ_2 ^3$  \\ 
& \varstr{17pt}{14pt} $({{21, \{23, 25, 27, 28\}}})$ & $\IZ_4 ^2 {\times} \IZ_2$ & $\IZ_2 {\times} (\IZ_4 ^2 {\rtimes} \IZ_2)$ & $\ID_8$ \\ 
& \varstr{17pt}{14pt} \begin{minipage}[c][80pt][c]{1.5in}
\begin{gather*}
(21, 29-30), (2568, 41\\
-42), (6836, \{96, 98, 99,\\
100, 102, 104-106\}),\\
(7735, 6-7), (7861, 6)  \\ 
\end{gather*}
\end{minipage} & $\IZ_4 {\times} \IZ_2$ & $\IZ_2 {\times} \ID_8$ & $\IZ_2$ \\ 
& \varstr{17pt}{14pt} \begin{minipage}[c][50pt][c]{1.5in}
\begin{gather*}
(6836, \{94, 107,\\
108, 110\}) \\ 
\end{gather*}
\end{minipage} & $\IZ_4 {\times} \IZ_2 ^2$ & $\IZ_2 {\times} ((\IZ_4 {\times} \IZ_2) {\rtimes} \IZ_2)$ & $\IZ_2 ^2$ \\ \hline
\varstr{17pt}{14pt} $\IZ_2 ^3$ & $\mathbf{({{7861, {8}}})}$ & $\IZ_2 ^ 7$ & $\IZ_2 ^ 7$ & $\IZ_2 ^ 4$ \\ \hline
\multirow{4}{*}{\raisebox{-1.8cm}{$\IQ_8$}} & \varstr{17pt}{14pt} \begin{minipage}[c][70pt][c]{1.5in}
\begin{gather*}
(19, \{21, 22, 24, 25, \\
27, 28\}), (21, \{33, 34\}),\\
 (2564, 7-9)\\ 
\end{gather*}
\end{minipage} & $\IZ_4$ & $(\IZ_4 {\times} \IZ_2) {\rtimes} \IZ_2$ & $\IZ_2$ \\ 
&\varstr{17pt}{14pt} $({{19, \{23, 26, 29\}}})$ & $\IZ_4$ & $\IZ_4 ^2 {\rtimes} \IZ_2$ & $\IZ_4$\\ 
&\varstr{17pt}{14pt} $({{21, {32}}})$ & $\IQ_8$ & $(\IZ_4 ^2 {\rtimes} \IZ_2) {\rtimes} \IZ_2$ & $\ID_8$  \\ 
&\varstr{17pt}{14pt} $({{6836, {112-113}}})$ & $\IQ_8$ & $(\IZ_2 {\times} \ID_8) {\rtimes} \IZ_2$ & $\IZ_2 ^2$  \\ \hline
\varstr{17pt}{14pt} \multirow{3}{*}{\raisebox{-1.0cm}{$\IZ_3{\times}\IZ_3$}} & $({{14, {4-7}}})$ & $\IZ_3 ^4$ & $\IZ_3 ^ 4 {\rtimes} \IZ_2$ & $\IZ_3 ^ 2 {\rtimes} \IZ_2$  \\ 
\varstr{17pt}{14pt} &  $({{14, {8-39}}})$ & $\IZ_3{\times}\IZ_3$ & $\IZ_3 ^3$ & $\IZ_3$\\
\varstr{17pt}{14pt}&  $({{7878, {2-3}}})$ & $\IZ_3 ^2$ & $\IZ_3 ^ 2 {\rtimes} \IZ_2$ & $\IZ_2$ \\ \hline
\varstr{17pt}{14pt}  $\IZ_{10}$ & $({{7447, {4}}})$ & $\IZ_{10}$ & $\ID_{20}$ & $\IZ_2$ \\ \hline
\varstr{17pt}{14pt} $\IZ_3 {\rtimes} \IZ_4$ & $({{7246, {21-23}}})$ & $\IZ_2$ & $(\IZ_6 {\times} \IZ_2) {\rtimes} \IZ_2$ & $\IZ_2$  \\ \hline
\multirow{3}{*}{\raisebox{-1.1cm}{$\IZ_4{\times}\IZ_4$}} & \varstr{17pt}{14pt} $({{21, {35-37}}})$ &  $\IZ_4 ^2$ & $(\IZ_2 ^3 {\times} \ID_8) {\rtimes} \IZ_2$ & $\IZ_2 ^3$ \\ 
&\varstr{17pt}{14pt} $({{7861, \{9, 10\}}), ({7862, {7}}})$ & $\IZ_4 ^2$ & $\IZ_4 ^2 {\rtimes} \IZ_2$ & $\IZ_2$  \\ 
&\varstr{17pt}{14pt} $({{7861, {11}}})$ & $\IZ_4 ^2 {\times} \IZ_2$ & $\IZ_2 {\times} (\IZ_4 ^2 {\rtimes} \IZ_2)$ & $\IZ_2 ^2$  \\ \hline
\multirow{3}{*}{\raisebox{0.7cm}{$\IZ_4{\rtimes}\IZ_4$}} & \varstr{17pt}{14pt} $({{21, {38-40}}})$ & $\IZ_4 {\times} \IZ_2$ & $(\IZ_2 ^3 {\times} \ID_8) {\rtimes} \IZ_2$ & $\IZ_2 ^3$  \\ 
&\varstr{17pt}{14pt} $({{6836, {114-115}}})$ & $\IZ_2 ^2$ & $(\IZ_2 ^2 {\times} \ID_8) {\rtimes} \IZ_2$ & $\IZ_2 ^2$ \\ 
&\varstr{17pt}{14pt} $({{7861, {12}}), ({7862, {8}}})$ & $\IZ_2 ^2$ & $(\IZ_4 {\times} \IZ_2 ^2) {\rtimes} \IZ_2$ & $\IZ_2$  \\ \hline
\multirow{3}{*}{\raisebox{-1.9cm}{$\IZ_8{\times}\IZ_2$}} & \varstr{17pt}{14pt} $({{21, {41}}})$ & $\IZ_8 {\times} \IZ_2$ & $(\IZ_2 ^3 {\times} \ID_8) {\rtimes} \IZ_2$ & $\IZ_2 ^3$  \\ 
&\varstr{17pt}{14pt}  $({{21, {42-43}}})$ & $\IZ_8 {\times} \IZ_2$ & $\IZ_2 {\times} ((\IZ_2 {\times} \ID_8) {\rtimes} \IZ_2)$ & $\IZ_2 ^2$  \\ 
&\varstr{17pt}{14pt} \begin{minipage}[c][50pt][c]{1.5in}
\begin{gather*}
(6836, 116-117),\\
(7861, 13) \\ 
\end{gather*}
\end{minipage} & $\IZ_8 {\times} \IZ_2$ & $\IZ_2 {\times} (\IZ_8 {\rtimes} \IZ_2)$ & $\IZ_2$ \\ 
&\varstr{17pt}{14pt} $({{7862, {9}}})$ & $\IZ_8 {\times} \IZ_2$ & $\IZ_2 {\times} \ID_{16}$ & $\IZ_2$  \\ \hline
\multirow{2}{*}{\raisebox{-0.8cm}{$\IZ_8 {\rtimes} \IZ_2$}} &\varstr{17pt}{14pt} $({{21, {44-45}}})$  & $\IZ_4 {\times} \IZ_2$ & $\IZ_2 {\times} ((\IZ_2 {\times} \ID_8) {\rtimes} \IZ_2)$ & $\IZ_2 ^2$   \\ 
&\varstr{17pt}{14pt} $({{21, {46}}})$ & $\IZ_4 {\times} \IZ_2$ &
\begin{minipage}[c][50pt][c]{1.5in} \begin{gather*} 
(\IZ_2 {\times} ((\IZ_2 {\times} \ID_8) {\rtimes} \IZ_2)) \\
{\rtimes} \IZ_2 \\
\end{gather*}
\end{minipage} & $\ID_8$   \\ \hline
\multirow{6}{*}{\raisebox{-3.2cm}{$\IZ_2 {\times} \IQ_8$}}&\varstr{17pt}{14pt} $({{21, \{47, 48, 50\}}})$ & $\IZ_2 {\times} \IQ_8$ & $\IZ_2 {\times} ((\IZ_4 ^2 {\rtimes} \IZ_2) {\rtimes} \IZ_2)$ & $\ID_8$   \\ 
&\varstr{17pt}{14pt} $({{21, {49}}})$ & $\IZ_2 {\times} \IQ_8$ & $(\IZ_2 ^2 {\times} (\IZ_4 ^2 {\rtimes} \IZ_2)) {\rtimes} \IZ_2$ & $\IZ_2 {\times} \ID_8$  \\ 
&\varstr{17pt}{14pt} $({{21, {51}}})$ & $\IZ_2 {\times} \IQ_8$ & 
\begin{minipage}[c][50pt][c]{1.5in} \begin{gather*} 
\IZ_2 {\times}(((\IZ_2 {\times} \ID_8) {\rtimes} \IZ_2)\\
 {\rtimes} \IZ_2)\\
\end{gather*}
\end{minipage}
& $\ID_8$  \\ 
&\varstr{17pt}{14pt} $({{21, {52-53}}})$ & $\IZ_2 {\times} \IQ_8$ & $(\IZ_2 ^3 {\times} \ID_8) {\rtimes} \IZ_2$ & $\IZ_2 ^3$  \\ 
&\varstr{17pt}{14pt} $({{7861, {17-19}}})$ & $\IZ_4 {\times} \IZ_2$ & $\IZ_2 {\times} ((\IZ_4 {\times} \IZ_2) {\rtimes} \IZ_2)$ & $\IZ_2$  \\ 
&\varstr{17pt}{14pt}  $({{7862, {11}}})$  & $\IZ_2 {\times} \IQ_8$ & $\IZ_2 {\times} ((\IZ_2 {\times} \ID_8) {\rtimes} \IZ_2)$ & $\IZ_2 ^2$  \\ \hline
\varstr{17pt}{14pt} $\IZ_4 {\times} \IZ_2 ^2$ & $({{7861, {14-16}}})$ & $\IZ_4 {\times} \IZ_2 ^3$ & $\IZ_2 ^2 {\times} (\IZ_2 ^ 4 {\rtimes} \IZ_2)$ & $\IZ_2 ^3$  \\ \hline
\varstr{17pt}{14pt} $\IZ_{10} {\times} \IZ_2$ & $({{7447, {5}}})$ & $\IZ_{10} {\times} \IZ_2$ & $\IZ_2 ^2 {\times} \ID_{10}$ & $\IZ_2$  \\ \hline
\varstr{17pt}{14pt} $\IZ_5 {\times} \IZ_5$ & $({{7890, {2-5}}})$  & $\IZ_5 {\times} \IZ_5$ & $(\IZ_5 {\times} \IZ_5) {\rtimes} \IZ_2$ & $\IZ_2$ \\ \hline
\multirow{2}{*}{\raisebox{0.5cm}{$(\IZ_4 {\times} \IZ_2) {\rtimes} \IZ_4$}} & \varstr{17pt}{14pt} $({{7861, {20}}})$ & $(\IZ_4 {\times} \IZ_2) {\rtimes} \IZ_4$ & $(\IZ_2 ^2 {\times} (\IZ_2 ^ 4 {\rtimes} \IZ_2)) {\rtimes} \IZ_2$ & $\IZ_2 ^3$ \\ 
& \varstr{17pt}{14pt} $({{7861, {21-23}}})$ & $\IZ_2 ^3$ & 
\begin{minipage}[c][50pt][c]{1.5in} \begin{gather*} 
(\IZ_2 ^2 {\times} ((\IZ_2 {\times} \ID_8) {\rtimes} \IZ_2)) \\
{\rtimes} \IZ_2\\
\end{gather*}
\end{minipage} & $\IZ_2 ^3$ \\ \hline
\varstr{17pt}{14pt}  $\IZ_8 {\times} \IZ_4$ & $({{7861, {24-25}}})$ & $\IZ_8 {\times} \IZ_4$ & 
\begin{minipage}[c][50pt][c]{1.5in} \begin{gather*} 
(\IZ_2 {\times} ((\IZ_2 {\times} \ID_8) {\rtimes} \IZ_2))\\
 {\rtimes} \IZ_2\\
\end{gather*}\end{minipage}
 & $\IZ_2 ^2$  \\ \hline
$\IZ_8 {\rtimes} \IZ_4$ & \varstr{17pt}{14pt} $({{7861, {26}}})$\footnote{The two distinct semi-direct products  $\IZ_8{\rtimes}\IZ_4$ correspond to the presentations $\langle a,b \mid a^8 = b^4 = e, bab^{-1} = a^3 \rangle$ and $\langle a,b \mid a^8 = b^4 = e, bab^{-1} = a^5 \rangle$.\label{semidirect}} & $\IZ_4 ^2$ & 
\begin{minipage}[c][50pt][c]{1.5in} \begin{gather*} 
(\IZ_2 {\times} ((\IZ_2 {\times} \ID_8) {\rtimes} \IZ_2))\\
 {\rtimes} \IZ_2\\
\end{gather*}\end{minipage} & $\IZ_2 ^2$  \\
{ $\IZ_8 {\rtimes} \IZ_4$} &\varstr{17pt}{14pt} $({{7861, {28}}})$\footref{semidirect} & $\IZ_2 ^2$ & $(\IZ_2 {\times} \ID_{16}) {\rtimes} \IZ_2$ & $\IZ_2$  \\ \hline
\varstr{17pt}{14pt}  $(\IZ_8 {\times} \IZ_2) {\rtimes} \IZ_2$ & $({{7861, {27}}})$ & $\IZ_4 {\times} \IZ_2$ & 
\begin{minipage}[c][50pt][c]{1.5in} \begin{gather*} 
(((\IZ_8 {\times} \IZ_2) {\rtimes} \IZ_2) {\rtimes} \IZ_2)\\
{\rtimes} \IZ_2 \\
\end{gather*}
\end{minipage} & $\IZ_2 ^2$   \\ \hline
\varstr{17pt}{14pt}  $\IZ_4 ^2 {\times} \IZ_2$ & $({{7861, {29-36}}})$ & $\IZ_4 ^2 {\times} \IZ_2$ & $(\IZ_2 ^2 {\times} (\IZ_2 ^ 4 {\rtimes} \IZ_2)) {\rtimes} \IZ_2$ & $\IZ_2 ^3$  \\ \hline
\varstr{17pt}{14pt}   $\IZ_4 {\rtimes} \IQ_8$ & $({{7861, {39}}})$ & $\IZ_4 {\times} \IZ_2$ & 
\begin{minipage}[c][50pt][c]{1.5in} \begin{gather*} 
(\IZ_2 {\times} ((\IZ_2 {\times} \ID_8) {\rtimes} \IZ_2))\\ {\rtimes} \IZ_2\\
\end{gather*}\end{minipage} & $\IZ_2 ^2$  \\ \hline
\varstr{17pt}{14pt} $\IZ_2{\times}\!(\IZ_4{\rtimes}\IZ_4)$ & $({{7861, {37-38}}})$ & $\IZ_2 ^3$ & $(\IZ_2 ^2 {\times} (\IZ_2 ^ 4 {\rtimes} \IZ_2)) {\rtimes} \IZ_2$ & $\IZ_2 ^3$   \\ \hline
\varstr{17pt}{14pt} $\IZ_2 ^2 {\times} \IQ_8$ & $({{7861, {40-45}}})$ & $\IZ_2 ^2 {\times} \IQ_8$ & 
\begin{minipage}[c][50pt][c]{1.5in} \begin{gather*} 
(\IZ_2 ^2 {\times} ((\IZ_2 {\times} \ID_8) {\rtimes} \IZ_2))\\ {\rtimes} \IZ_2\\
\end{gather*}
\end{minipage} & $\IZ_2 ^3$  \\ 
\hline
\end{longtable}
\end{center}

\end{appendix}
\newpage

\end{document}